\documentclass[12pt]{elsarticle}
\usepackage{graphicx}
\usepackage{color}
\usepackage{latexsym}
\usepackage{amsmath}
\usepackage{amssymb}

\newcommand{\nc}{\newcommand}

\nc{\eps}{\varepsilon}
\nc{\vp}{\varphi}
\nc{\bea}{\begin{eqnarray}}  \nc{\eea}{\end{eqnarray}}
\nc{\beq}{\begin{equation}}  \nc{\eeq}{\end{equation}}
\nc{\ben}{\begin{enumerate}} \nc{\een}{\end{enumerate}}

\nc{\tvp}{\widetilde{\varphi}}
\nc{\D}{\mbox{$\not\!\!D$}}
\nc{\Db}{\mbox{${\raisebox{2mm}{\boldmath ${}^\leftarrow$}\hspace{-4mm} D}$}}
\nc{\Dbs}{\mbox{${\raisebox{2mm}{\boldmath ${}^\leftarrow$}\hspace{-4mm} \D}$}}
\nc{\Dfb}{\mbox{$\raisebox{2mm}{\boldmath ${}^\leftrightarrow$}\hspace{-4mm} D$}}
\nc{\Dfbs}{\mbox{$\raisebox{2mm}{\boldmath ${}^\leftrightarrow$}\hspace{-4mm} \D$}}
\nc{\vpj }{\mbox{${\vp^\dag i\,\raisebox{2mm}{\boldmath ${}^\leftrightarrow$}\hspace{-4mm} D_\mu\,\vp}$}}
\nc{\vpjt}{\mbox{${\vp^\dag i\,\raisebox{2mm}{\boldmath ${}^\leftrightarrow$}\hspace{-4mm} D_\mu^{\,I}\,\vp}$}}
\nc{\psid}{\mbox{$\overline{\psi} i\,\raisebox{2mm}{\boldmath ${}^\leftrightarrow$}\hspace{-4mm} \D\,\psi$}}
\nc{\f}{\frac}

\def\gev{\hbox{GeV}}
\def\vev{vacuum expectation value}
\def\eg{{\it eg.}}
\def\lcal{{\cal L}}
\def\vevof#1{\left\langle #1 \right\rangle}
\def\inv#1{\frac1{#1}}
\def\up#1{^{\left( #1 \right) }}
\def\half{\frac12}
\def\lesim{\lesssim}
\def\tev{\hbox{TeV}}
\def\sm{Standard Model}
\def\su#1{{SU(#1)}}

\def\wt{\widetilde}

\def\ocal{{\cal O}}

\def\etal{{\it et al.}} 

\newcommand{\eqn}[1]{eq.~(\ref{#1})}
\newcommand{\eqns}[2]{eqs.~(\ref{#1}),(\ref{#2})}

\def\qwe{f} %effective operator coefficient
\def\mn{{\mu\nu}}

\journal{Nuclear Physics B}

\begin{document}

\begin{frontmatter}



\title{%
%{%
%\vspace{-5.0cm}
%\small\hfill\parbox{32.0mm}{\raggedleft%
%MCTP-13-23\\
%NSF-KITP-13-145\\
%UCRHEP-T532\\
%}}\\[.65cm]
Higgs-Boson Couplings Beyond the Standard Model}


\author[label1,label2]{Martin B Einhorn}
\author[label3]{Jos\'e Wudka}
\address[label1]{Kavli Institute for Theoretical Physics,University of California, Santa Barbara, CA 93106-4030\footnote{Current address.}}
\address[label2]{Michigan Center for Theoretical Physics, University of Michigan, Ann Arbor, MI 48109}
\address[label3]{Department of Physics and Astronomy, University of California, Riverside, CA 92521-0413}

\begin{abstract}
The implications for Higgs decays of potential new physics
beyond the Standard Model (BSM) are considered in the context of
effective field theory, assuming perturbative decoupling. 
Using existing data to restrict which
dimension-six operators can arise, it is shown that, given the
existing experimental constraints, only a small number of
operators can affect the decays of the Higgs: those
that may be potentially-tree-generated (PTG) 
and modify the Higgs-fermion couplings, 
or those that may be loop-generated (LG) 
that modify the Higgs couplings 
to $\gamma\gamma,~Z\gamma$ and $GG$. 
Implications for specific branching ratios are given in terms of the coefficients of various dimension-six operators. In such a scenario, the ratios $ \Gamma\left(H\to W W^*\right)/ \Gamma\left(H\to ZZ^*\right)$
and $ \Gamma\left(H\to W \ell\nu\right)/ \Gamma\left(H\to Z\ell\ell\right)$ equal to their standard model values to an accuracy 
of  $O(1\%)$ or less.

\end{abstract}


\end{frontmatter}

 %\linenumbers


\section{Introduction} \label{sec:intro}

The observation of a new particle by two detectors
\cite{:2012gk,:2012gu} at the LHC has offered a candidate for
the long-sought Higgs boson of the Standard Model (SM). Within
errors, the observed properties are consistent with the Higgs
boson of the SM insofar as its production rate and branching
ratios are concerned, coming primarily from data in the $WW^*,$
$ZZ^*$, and $\gamma\gamma$ decay modes. There is some supporting
evidence from enhancements in TeVatron data \cite{:2012zzl},
primarily from $b\bar{b}$ decays. So far, the evidence is
consistent with SM expectations, with the exception of the rate
in the $\gamma\gamma$-channel, which apparently exceeds SM
estimates in one of the experiments \cite{:2012gk}.

Naturally, a primary goal of further experiments is to determine
whether the couplings of the Higgs to weak bosons $g_{HWW},$
$g_{HZZ}$ and the couplings to fermions agree with the SM
expectations. These studies may be informed by theoretical
expectations, and many papers have been
written (for recent summaries see e.g. \cite{h-eff})
about the implications of models that include particles
beyond the SM (BSM). With a mass $m_H\approx 125~\gev,$ the
Higgs would appear to have a small enough self-coupling 
for perturbation theory to be reliable.
In that case, there are two possibilities
for the additional particles in such models: either (1)~they
involve new ``light" particles of a mass comparable to or lighter than
the Higgs boson, as, for example, in models having two Higgs
doublets, including some supersymmetric models\footnote{Some
recent fits of such models to LHC data are
\cite{Ferreira:2012nv, Belanger:2012gc}.}, 
in which most or all of their mass derives from the electroweak scale as do SM particles, or (2)~all new
particles are more massive than $m_H,$ deriving their mass from some new underlying scale. An example would be softly-broken supersymmetric models with super-renormalizable couplings large compared to the weak scale. In the absence of the observation of a new particle, it can be difficult to decide in which situation we find ourselves.

The question is, what can be inferred from deviations of experimental data from SM expectations of the properties of the observed scalar?  In the former case, there tend to be rather large deviations from the SM, arising from mixing angles between two or more multiplets.  Typically, couplings of a Higgs boson differ already at tree level by factors such as $\tan\beta,$ the ratio of \vev s of different doublets.  In the latter case, one
may perform a model-independent analysis using a generic effective Lagrangian approach, taking into account that the first corrections to the SM can be described in terms of higher-dimensional operators (HDO). A large number of publications have appeared recently that discuss various aspects of this approach; it would require a lengthy review to cite all the papers that have been written on this, and such a list would be out of date by the time of this publication; for a  representative sample see~\cite{legion}.

A related question is, if there are no other particles discovered and no deviations from the SM observed, what conclusions can be drawn, given the level of accuracy of the experiments?  As much as possible, one would like to draw model-independent conclusions, although that may be very difficult in the near term.

Some years ago, the authors \cite{Arzt:1994gp} and others~\cite{De Rujula:1991se, Hagiwara:1993ck} performed such analyses, both
for weakly-interacting, decoupling scenarios and for strongly interacting models~\cite{Ellison:1998uy, Pich:2012dv}.
In this paper, we shall assume the underlying physics is decoupling
and weakly coupled, at least to a good approximation.  What this means in practice is that the particles that we call ``light," such as the top quark, predominantly get their masses as a result of spontaneous breaking of $SU(2)\otimes U(1).$  We assume that the ``heavy" particles get their masses primarily via some other mechanism, although they may also receive electroweak contributions.  For example, this would be the case if the scale of supersymmetry-breaking were large compared to the weak scale, giving some superpartners parametrically large masses. 

In phenomenological studies of deviations from the SM, it has been advocated\footnote{See, \eg, the reviews in refs.~\cite{Belanger:2012gc, ZwirnerMoriond}.} that, to fit experimental events involving the production and decay of a single Higgs boson, one employ an effective Lagrangian of the form
\bea\label{lone}
\lcal_{eff}\!\!\!&=&\!\!\!\frac{H}{v}\left[\left(2c_W M_W^2 W^-_\mu W^+_\mu +c_Z M_Z^2 Z_\mu^2\right)
+c_t m_t t\bar{t}+c_b m_b b\bar{b}+c_\tau m_\tau \tau\bar{\tau}\right] \\
&+&\frac{H}{3\pi v}\left[ c_\gamma \frac{2\alpha}{3}F_\mn^2+c_g\frac{\alpha_S}{4} G_\mn^2\right].
\label{ltwo}
\eea
This effective Lagrangian is presumed to describe interactions in the so-called unitary-gauge, where the Higgs doublet is of the form $ \phi = ((v+ H)/\sqrt{2}) (0,1)^T$ (and $ v = \sqrt{2} \vevof\phi \simeq 246~\gev $.)  In the SM, the interactions described in \eqn{ltwo} arise at the one-loop level, whereas those in \eqn{lone} arise already in tree approximation.   To leading order, all the coefficients $c_k=1,$  which, given present experimental accuracies, is sufficient, although further radiative corrections can be included if necessary.

Although the decay rates are unambiguous in the Standard Model, unless we know the form of the BSM Lagrangian, one cannot blithely use eqs.~(\ref{lone}) and (\ref{ltwo}) as an effective Lagrangian without acknowledging other implications of such a choice.
Because of the {\it  equivalence theorem,} which we will review in the next section, the form of an effective Lagrangian involves a certain degree of arbitrariness.  The particular choice of operators does not affect expectations for S-matrix elements, but the associated Green's functions may be very different.   At the very least, these additional assumptions need to be spelled out in detail.  The expressions \eqns{lone}{ltwo}, although 
not gauge-invariant, are intended to be used in unitary gauge. 
Some decays can be well approximated by treating all particles on-shell, so that, \eg, $c_b, c_\tau, c_\gamma$ may be defined with all three particles on-mass-shell, other parameters such as $c_W, c_Z, c_t$ represent coupling constants that, for kinematical reasons, cannot be determined experimentally. 
As an example, the $ H \to WW^* $ mode represents
$ H \to W \ell \nu_\ell $ (where $\ell=e,\mu$), which receives a contribution from a virtual $W$ exchange, but others as well,
and these depend on the operator basis being used. Extracting limits on coefficients such as $ c_W$ requires 
a complete calculation that includes all
 relevant contributions to insure the results are independent
of the operator basis and fully gauge invariant.

In fact, among other results below, we shall show that modifications to the SM couplings from new physics $c_W, c_Z$ are negligible
within foreseeable experimental errors (except for a possible common normalization effect that does not contribute to the branching ratios),
 unless there are other light particles whose masses arise primarily, if not solely, from electroweak symmetry-breaking.  
In that case, the form of \eqn{lone} is unsuitable as a starting point for fitting or interpreting experimental data.  No conclusions can be drawn from it without making presumptions about other operators and processes involving the Higgs boson.

%%%%%%%%%%%%%%%%%
\section{Some features of effective Lagrangians}

By now, the language of effective field theory has become familiar~\cite{Georgi:1994qn, Weinberg:1996kr, AlvarezGaume:2012zz, Zee:2003mt}, especially to researchers studying physics BSM.  We will review it here only to the extent that we need to establish notation and to summarize some features of the approach; the details of our general approach are provided in a companion paper~\cite{Einhorn:2013kja}.

We imagine a theory where the heavy scale $ \Lambda $ is assumed too large for the corresponding excitations to be directly produced; their virtual effects, however, may be observable. Assuming also that the heavy physics decouples then implies that at scales below $ \Lambda $ the effective action can be expanded in a power series in $ \Lambda $ (multiplied by logarithmic corrections) where all each power multiplies a local operator, and those terms that grow with $ \Lambda $ can be absorbed in a renormalization of the low-energy parameters. After this renormalization the effective action takes the form
\beq
S_{\rm eff} = \int d^4x \lcal_{\rm eff} \quad
\lcal_{\rm eff} = \lcal_0 + \sum_{n \ge 5} \inv{\Lambda^n}
\sum_i \qwe\up n_i \ocal\up n_i,
\eeq
where we assumed the underlying physics is weakly coupled\footnote{Although $\lcal_{\rm eff}$ must be Hermitian, it is not always most expedient to make each term in the sum Hermitian; \eg, in the SM, the Yukawa couplings are an illustration. In such cases, each term is implicitly
accompanied by its Hermitian conjugate.}. The coefficients $\qwe_i $ encode all the details of the underlying model and can therefore be used to parameterize all possible types of heavy physics; in general they can also depend logarithmically on $ \Lambda $.
For the case being considered here,  $ \lcal_0 $ corresponds to the full SM Lagrangian, in which case there is a single dimension-five~\cite{Weinberg:1979sa} operator  that violates lepton number by two units and generates a neutrino Majorana mass. Aside from this, dimension-six operators then represent the leading virtual new physics effects resulting from any weakly-coupled,  decoupling heavy particles.

The number of dimension-six operators is large $(\sim100)$, but not all need be included in calculations since all low-energy effects can be parametrized by the coefficients of a reduced set of operators we refer to as a basis. This is a set of (dimension-six) operators $ \{\ocal_a \}$ (henceforth we drop the superscript $(6)$) with the property that any other operator $ \ocal $ obeys the relation
\beq
\ocal - \sum_a \kappa_a \ocal_a = \sum_\phi U_\phi \frac{\delta S_0}{\delta \phi},
\label{eq:equiv}
\eeq
where the $ \kappa_a $ are appropriately chosen constants; $ \phi, $ a generic light field; ${\delta S_0}/{\delta \phi},$ the classical equations of motion; and $ U_\phi ,$ local operators.  Each term in \eqn{eq:equiv} is gauge- and Lorentz-invariant. We will say that the combination $ \ocal - \sum \kappa_a \ocal_a $ vanishes ``on-shell," and that $ \ocal$ is {\it equivalent} to $\sum \kappa_a \ocal_a$. In addition, we demand that no linear combination of basis elements vanishes on shell.

To establish some terminology, note that the HDO form a vector space.  An equivalence relation produces a unique partition of a vector space into equivalence classes.  A basis will have one operator from each equivalence class.  A minimal basis choice for dimension-six operators is presented in~\cite{Grzadkowski:2010es}. 
While the preceding is a familiar construction, in~\cite{Einhorn:2013kja}, we put forward an improved method of choosing basis operators with reference how these operators may arise in extensions of the SM.
  An {\it extension} of $\lcal_0$ is any model containing heavy particles that reduces to $\lcal_0$ for operators of dimension four or less at scales below some threshold $\Lambda.$  An extended model may also be referred to as an {\it embedding} of $\lcal_0.$

In general, HDO may be identified as either {\it poten\-tially-tree-generated} (PTG) or {\it loop-generated} (LG)~\cite{Arzt:1994gp,Einhorn:2013kja}.
A PTG operator $\ocal^{PTG}$ is one for which an extension can be found
in which it arises from a tree-diagram.  An LG operator $\ocal^{LG}$ is one that (1)~cannot emerge from a tree-graph in any embedding of 
$\lcal_0$ and (2)~can arise from loop-graphs\footnote{This classification can be made either without restriction on the embeddings of $\lcal_0$ or, if one is interested in a restricted set of embeddings, from extended models respecting some additional local or global symmetry.  For example, the limits on violation of baryon- or lepton-number suggest that such operators may be ignored for analyzing LHC data, regardless of whether they are PTG or LG.}.   This is a useful distinction because LG operators have coefficients that are typically suppressed by a factor $ \sim1/(4\pi)^{2n} $ relative to PTG operators, where $n$ is the number of loops.  Having potentially larger coefficients, the PTG operators may be more sensitive to new physics effects.

It is important to note that whether an operator is LG or PTG is a property of the heavy physics, while an equivalence relation of the form \eqn{eq:equiv} is a property of the light theory.  An equivalence class of operators may contain only PTG-operators, only LG-operators, or both kinds. It is helpful to identify this property of each equivalence class and, in cases when a class contains both kinds of operators, to choose a basis operator from among the PTG-operators.   
This provides the most conservative approach to interpreting experimental data, whether providing limits on or evidence for BSM physics effects.  The reason is that the parametrization covers the widest class of heavy physics theories: those that generate the operators in question at tree-level, as well as those that may generate them only through loop corrections.
In~\cite{Einhorn:2013kja}, we delineated the equivalence relations for the SM and analyzed the basis chosen in~\cite{Grzadkowski:2010es}.  We identified those that were LG, those that were PTG, and in cases where an equivalence class contained both types, showed that the basis chosen in~\cite{Grzadkowski:2010es} were in fact PTG operators, as required.  This then is good basis for studying physics BSM.  In the remainder of this paper, we indicate how this may be applied to Higgs production and decay.

%%%%%%%%%%%%%%%%%%%%%%%
\section{New physics contributions to Higgs decay}

Limiting our attention to baryon- and lepton-conserving operators
involving only SM fields (e.g. no right-handed neutrinos), and 
adopting the basis $ \{ \ocal_a \}$ of dimension-six operators given in ref~\cite{Grzadkowski:2010es}, the effective Lagrangian takes the form\footnote{Although we will use the basis of~\cite{Grzadkowski:2010es}, we prefer to denote the operators as $\ocal_a$ rather than $Q_a$.  For convenience, we reproduce them in \ref{sec:tables}.}
\beq
\lcal_{\rm eff} = \lcal_{SM} + \sum_a \frac{\qwe_a }{\Lambda^2} \ocal_a + \cdots,
\eeq
where the ellipsis denote operators of dimension $ > 6 $. We need only PTG operators that contribute to the various Higgs decay channels, except for the 
$ \gamma\gamma,$ $\gamma Z,$ and $GG$ final states which occur at one-loop in the SM and in all extensions of the SM.  These therefore, invite comparison with one-LG corrections to new physics.
  As discussed above, the operators included describe the leading deviation from the SM, whereas those we neglect will be too small to be observed, at least in present experiments. In the following we will assume that the
coefficients $\qwe_a$ are real\footnote{For those operators $\ocal_a$ that are Hermitian, $f_a$ are necessarily real; for others,  including the ones relevant in Higgs decays, it is equivalent to assuming that CP violation is unimportant for these applications.}.

We find that the PTG operators contributing to the Higgs decay channels measured at the LHC are also involved in other well-measured process, namely, $Z$ and $W$ lepton decays and custodial symmetry violations associated with the oblique $T$ (or $ \rho $) parameter. Current data indicate that deviation from the SM in these processes lie below the level of 0.1\%--1\% , so that the contributions to the corresponding operators to Higgs decays can be neglected given the current precision in that decay.

The PTG operator basis contributing to the measured Higgs decays can be separated into 3 classes:
\ben

\item Pure Higgs operators\footnote{In \cite{Grzadkowski:2010es}, $\ocal_{\partial\phi}$ is replaced by $-\ocal_{\phi\Box}$, which is the same after integration by parts.}:
\beq
\ocal_{\partial\phi} = \textstyle\half( \partial_\mu |\phi|^2 )^2, \qquad \ocal_\phi=|\phi|^6.
\eeq
\hskip0.25in Ignoring self-interactions of the scalar field, the effect of $\ocal_{\partial\phi}$ will be to change the normalization of the Higgs field after symmetry breaking.
Indeed, in unitary gauge, 
\beq\label{eq:unitary}
\phi = \frac{1}{\sqrt{2}}\begin{pmatrix} 0 \\ v+ h\end{pmatrix},
\end{equation}
(with $ v = \sqrt{2} \vevof\phi \simeq 246~\gev $,)
so we get
\beq
\lcal = \lcal_{SM} + \frac{\qwe_{\partial\phi}}{\Lambda^2} \ocal_{\partial\phi} + \cdots \approx \half (1 +   \epsilon \qwe_{\partial\phi} )(\partial h)^2+\cdots,
\eeq
where $\epsilon\equiv v^2/\Lambda^2.$  Thus,
the canonically normalized Higgs field will be
\beq
H = \sqrt{1 +  \qwe_{\partial\phi} \epsilon\;}\, h\approx
\left(1 +  \textstyle\half \qwe_{\partial\phi} \epsilon \right)h.
\label{eq:rescale}
\eeq
This modifies all processes involving a single Higgs boson in the same way, replacing the SM Higgs field $h$ by $H.$

\hskip7mm The operator $\ocal_\vp$ also modifies the parameters in the Higgs sector, such as the \vev\ $v$, but these effects can be absorbed in finite renormalizations of $\lcal_0$ with no observable effects.  In that case, $\ocal_\vp$ could only be distinguished through Higgs self-coupling effects~\cite{Bonnet:2011yx,Bonnet:2012nm} which, at present, are experimentally out of reach. In contrast the rescaling (\ref{eq:rescale}) does have (potential) observable consequences.

%%%%%%%
\medskip
\item Operators modifying Higgs couplings to $W$ and $Z$:

\smallskip
\hskip7mm These include basis operators of the type $X^2\vp^2,$ called $\ocal_{\vp X},$ $\ocal_{\vp \wt{X}},$ $\ocal_{\vp WB}$ and $\ocal_{\vp \wt{W}B}$ (see Appendix).  All of these are LG operators and will be neglected in first approximation.  That means that {\it the Higgs coupling to $ZZ$ and $WW$ may be assumed to be SM to within about\/ $0.1-\!1\%$.}  This important result is analogous to our earlier result concerning BSM corrections to triple-gauge-boson couplings~\cite{Einhorn:2013kja, Arzt:1994gp}.

\medskip

\hskip7mm The basis we employ also contains an operator
of the type 
$\vp^4D^2$:  $\ocal_{\phi D}\! \equiv\!
\big| \phi^\dagger D^\mu\phi \big|^2.$
This operator would generate a mass shift for the $Z$ and produce a change in the $\rho$-parameter~\cite{Einhorn:1981cy} or, equivalently, the so-called oblique $T$ parameter~\cite{Peskin:1991sw} from its SM value, specifically
\beq
\delta T = - \inv\alpha \epsilon \qwe_{\phi D},
\eeq
where, in obtaining this relation, we ignored the effective-operator contributions to the Fermi constant~\cite{SanchezColon:1998xg}. Current  experimental constraints~\footnote{See the review by Erler and Langacker in~\cite{Beringer:1900zz}.} give $|\delta T| \lesim 0.1$, implying  that, even though $\ocal_{\phi D}$
may affect Higgs decays (specifically, the $H \to ZZ^*$ mode\footnote{See comments on this decay mode in next section.},)
these effects will be too small to be observed, given the experimental precision achievable at the LHC.

%%%%%%%
\medskip
\item Higgs and Gauge Boson Couplings to Fermions:

\smallskip
\hskip7mm These include all the operators of the types~\cite{Grzadkowski:2010es}
$\psi^2\vp^2D$, called $(\ocal_{\vp \psi})_{pr}$ and $(\ocal_{\vp ud})_{pr}$, and $ \psi^2X\vp $, called $(\ocal_{\psi X})_{pr}$, for any fermion $\psi$, where $p, r$ are family indices.  The former are all PTG operators, but the latter are all $LG.$  Therefore, in first approximation, we will neglect $\ocal_{\psi X}.$

\hskip7mm Limits on flavor-changing neutral currents\footnote{See the review by Ceccucci, Ligeti, and Sakai in~\cite{Beringer:1900zz}.} suggest that the thresholds for generation-changing operators $(\ocal_{\vp \psi })_{pr}$ in the coupling of the $Z$ are very high, so we may assume that $p=r,$ but this still leaves the $Z$ couplings to the 3 families, $(\ocal_{\vp \psi })_{pp}$ for each type of fermion $\psi =\{\ell, e, q, u, d\}$.  Many of these are already precisely determined~\cite{Beringer:1900zz}, especially for $\psi=$~leptons for all 3 generations, from LEP measurements. 
The $Z$-lepton  couplings are measured to at least 1\%
and agree with the SM predictions to that precision, so their potential 
contributions to Higgs decay widths will lie in this range. 
These effects are of order $ \qwe \epsilon $ and correspond
to a scale $ \Lambda > 2.5 ~\tev$ when $ \qwe\sim 1 $.
For quarks, they are similar experimentally constrained for all flavors except for the $t$-quark. 
Although these operators break custodial symmetry and therefore would change 
$ | \delta T | $, their lowest order contributions are in one-loop corrections and so are not strongly constrained by the experimental value of $ | T | $.  To improve on existing limits significantly seems to be beyond the reach of a hadron collider such as LHC. 

\hskip7mm The operator $(\ocal_{\vp ud})_{pr}$ would modify couplings of the 
$W$ in ways that would affect both family-changing couplings as well as Higgs decays.  These couplings of the $W$ to the first two generations have been well-studied, and there are even constraints from $t \to b W.$  Once again, this would also contribute to $ | \delta T | $, but only at one loop order, and, given that the top quark only makes a contribution of about 0.2, a bound on the order of 10\% gives no significant constraint on the corresponding $\epsilon f_{\vp ud}$.
The operator $\ocal\up3_{\vp \ell}$ potentially modifies the $W\ell\nu$ coupling, but since measurements agree with the
\sm\ to a precision below 1\%, the effective operator modifications
to these couplings should lie in this range; as with the $Z$ case
their contributions to Higgs decays can be ignored.

\hskip7mm This case also includes corrections to the Yukawa couplings of the Higgs, which would affect the fermion masses as well as Higgs decays
\beq
(\ocal_{e\vp})_{pr}\! =\! |\vp|^2 \bar\ell_p e_r \vp, \ \ 
(\ocal_{u\vp})_{pr}\! =\! |\vp|^2 \bar q_p u_r \tilde\vp, \ \ 
(\ocal_{d\vp})_{pr}\! =\!  |\vp|^2 \bar q_p d_r \vp,
\eeq
where the fermions carry generational indices. If one simply replaces the Higgs field by its \vev, the contributions of these operators cannot be distinguished from the SM contribution.   Therefore, the $GIM$ mechanism will continue to work, as well as tests of CKM unitarity\footnote{See the review by Blucher and Marciano in~\cite{Beringer:1900zz}.}.  
However, they affect the Higgs decays to fermion pairs differently than SM Yukawa couplings, since these are cubic in the scalar doublet whereas Yukawa couplings are linear.  Thus, comparing the Higgs decay rates to the fermion masses may provide a good test for the presence of these operators.  At present, it has not yet been determined whether the SM couplings to the observed Higgs are proportional to the mass, so there is no constraint on the operator coefficients $\qwe_{\psi\vp}$.

\hskip7mm  Even with much increased precision, the projected limits on the effective operator coefficients may be harder to interpret (e.g. extracting limits on the scale of new physics) than one might think, depending on the nature of the underlying theory. In the SM, the vanishing of a fermion mass give rise to an enhanced chiral symmetry.  If that were a property of the HDO  as well, as is commonly believed to be the case,  then the coefficient of these operators ought to be proportional to at least the first power of the corresponding Yukawa coupling.  Thus, even fairly inaccurate measurements that limit the size of these kinds of corrections to Higgs decay would be a good indicator of whether this hypothesis is correct. 

\item Loop-generated operators:

In most applications, operators that are necessarily generated
by heavy-particle loops are disregarded as being subdominant,
or because their effects are expected to be difficult to detect experimentally.
The rare decays $ H \to \gamma\gamma,\, Z\gamma,\, GG $ (where $G$
represents a gluon) present exceptions since the
SM contributions are themselves loop-generated~\cite{Ellis:1975ap, Wilczek:1977zn}. In particular, the experimental
precision for $H \to  \gamma\gamma $ may eventually reach the level needed to detect or set limits on deviations generated by the heavy physics. These deviations are 
generated by {\it(i)} modifications of the vertices involved
in the SM loops (such as the $Ht\bar t $ coupling) and/or {\it(ii)}
contributions from operators generated by heavy-physics loops. 
The former are listed above, the latter are the following:
\beq
\ocal_{\phi X} = \half |\phi|^2 X_\mn X^\mn,~X=\{G^A,\, W^I,\,B\};
\quad
\ocal_{WB} = (\phi^\dagger \tau^I \phi)B^\mn W^I_\mn.
\label{eq:lg.ops}
\eeq
There are also operators involving the dual tensors $ \tilde X_\mn, $
but these do not interfere with the SM amplitudes, and so generate
contributions of order $ \Lambda^{-4} $ or smaller.
The coefficients of these operators are suppressed by a
loop factor $ 1/(16\pi^2) $; each field $X$ is also accompanied
by the corresponding gauge coupling constant since gauge fields couple
universally.

\een

Summarizing: in the following, given the precision anticipated for LHC experiments studying Higgs decays, we will ignore potential effects of the operator~$\ocal_{\vp D}$ and those of type $\psi^2\vp^2D$, {\it viz.}, $\ocal_{\vp\psi}$ and $\ocal_{\vp ud}$. 
 It is worth noting that these same operators
generate ``contact'' vertices of the form $HZee$ and $HWe\nu$ that could contribute to $H\to WW^*,~ZZ^*$ decays, so these too can be ignored.

\hskip7mm Thus, the only dimension-six operators to be retained for Higgs decays at LHC are $\ocal_{\partial\vp}$ and $\ocal_{\psi\vp}.$
 Recall that in the SM, the same transformation that diagonalizes the fermion masses will diagonalize the Higgs couplings, so there remain no (quark) flavor-changing couplings. When the  
 operators $(\ocal_{\psi\vp})_{pr}$ are introduced however, the diagonalization of the fermion mass matrix will not, in general, remove the flavor-changing Higgs couplings. The corresponding amplitudes, however, will be suppressed by a factor of $ \epsilon $, 
and therefore, the decay rates by $\epsilon^2.$  Concerning the flavor-diagonal operators, we will assume that their coefficients are proportional to the corresponding fermion masses, so that quantities of the form $ f_{\psi\vp}/m_\psi $ have a finite limit as $ m_\psi \to 0 $.  
This naturality assumption also makes these effects much harder to observe.

\section{Implications for LHC experiments}

We now determine the manner in which new heavy, decoupling physics can affect these decays while taking into consideration the limitations from existing data, as described above.

The operators $(\ocal_{\psi\vp})_{pp}, ~\psi=e_p,u_p,d_p $, together with $\ocal_{\partial\phi} ,$ modify the determination of the Yukawa couplings, which are not yet much experimentally constrained, as well as the Higgs decays into  fermions as follows:
\beq
\Gamma(H \to \bar\psi \psi) = \kappa_\psi^2\Gamma_{SM}(H \to \bar\psi \psi)
\,; \quad \kappa_\psi^2 =  \left( 1 -  \qwe_{\partial\phi} \epsilon + \frac{\sqrt{2}\, v}{m_\psi} \qwe_{\psi\vp}\epsilon\right).
\eeq

Before proceeding further, some comments are in order concerning the decays referred to by the CMS and ATLAS collaborations as $H \to Z Z^*$ and $H \to W W^*$. 
These are a shorthand for the average of the leptonic final states in which the virtual $Z^*$ decays into either $e^+ e^-$ or $\mu^+\mu^-$ and in which the $W^*$ decays into either $e\,\nu_e$ or $\mu\,\nu_\mu.$  If we consider the actual S-matrix elements for these processes, we find, \eg,  the decay to a $Z$ plus a lepton pair 
receives contributions from 3 diagrams:
$$ \includegraphics*[width=5in]{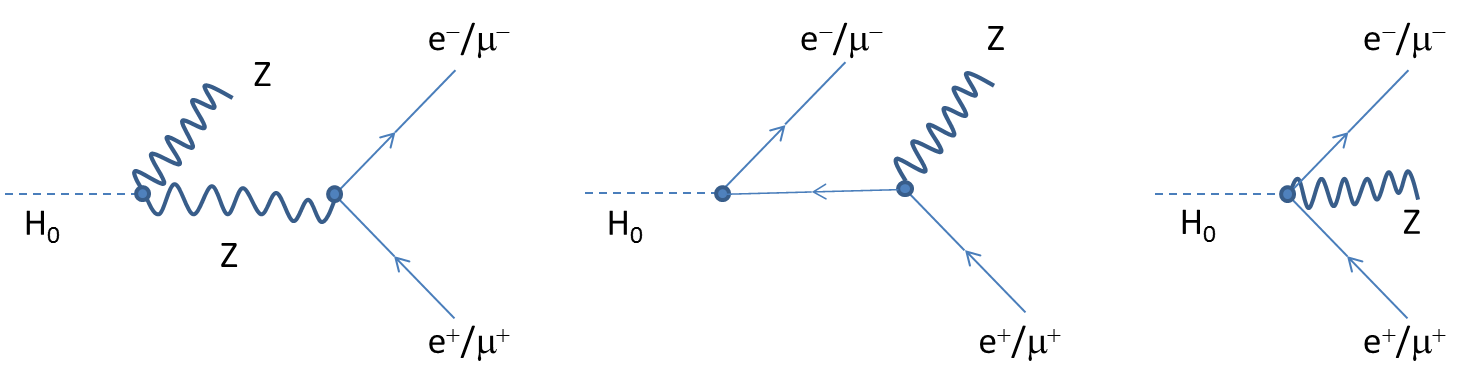} $$

The vertices here are intended to include the sum of the SM couplings and the corresponding PTG dimension-six operators.
 The first has a $Z$ internal line and, in addition to the SM contributions, is affected by $ \ocal_{\partial\phi}$. 
  The third graph involves a contact $HZee$ or $HZ\mu\mu$ interaction, which, we have argued in the previous section, can be neglected. 

The second diagram has an $e$ or $ \mu $ internal line.  As emphasized earlier, the first two diagrams are not gauge invariant in general, not even in the SM.  However, that gauge dependence is associated with the nonzero fermion mass, and the Yukawa couplings of the $e$ and $\mu$ make a tiny contribution to these decays.  We further assumed that the coefficients $\qwe_{\psi\vp}$ were proportional to the fermion mass.  If we set the mass zero, then the second diagram vanishes, and the first becomes gauge-invariant.  Stated otherwise, in the limit of vanishing fermion mass, both the vector and the axial-vector currents are conserved (the latter in the Goldstone mode.)  

Whether the preceding arguments remain true for the $\tau$-lepton remains to be seen, but the decay $Z\to\tau\tau$ agrees with the SM to the same accuracy as for decays to $ee$ and $\mu\mu.$

For whatever reasons, whether because of this approximate chiral symmetry or because their threshold $\Lambda$ is very large, the HDO's that could give rise to $\delta T$  give negligible corrections as well.

As a result, only  the effects from
$ \ocal_{\partial\phi}$  remain unconstrained, so we obtain
\beq
\Gamma(H \to ZZ^* ) = \kappa_Z^2 \Gamma_{SM}(H \to ZZ^* )
\,; \qquad \kappa_Z^2 =  \left( 1 - \qwe_{\partial\phi} \epsilon \right).
\eeq

A similar discussion applies to the $H\to WW^*$ mode: the effects generated
by possible deviations form the SM in the $W\ell\nu$ couplings
are well below the current experimental precision to which this decay mode
is measured.   So we find
\beq
\Gamma(H \to WW^* ) =  \kappa_W^2\Gamma_{SM}(H \to WW^* )
\,; \qquad \kappa_W^2 = \left( 1 - \qwe_{\partial\phi} \epsilon \right).
\eeq
The expected modification for both of these decay widths are the same because the contributions from $ \ocal_{\partial\phi} $ respect custodial symmetry,  so the ratio $ \Gamma(H \to ZZ^*)/\Gamma(H\to WW^*)$ equals their \sm\ value to an accuracy of 1\% or less.  If, on the contrary, this ratio is observed to differ markedly from 1, 
it is likely there will be other particles whose mass scale
are also generated primarily by electroweak symmetry-breaking, as occur in
models with more than one Higgs doublet and in supersymmetric models.

\medskip

Finally, we consider briefly three rare but important decays\footnote{
In obtaining the numbers below, we will substitute the following values 
for the top and Higgs masses and for the SM\ \vev:
$ m_t=173.5\,\gev,\, m_H=125\,\gev,\, v=246.22\,\gev $.}.
\ben
\item The $H \to  \gamma\gamma $ mode receives contributions both from tree-level modifications to the
$Htt$ and $HWW$ couplings, as well as from the loop-induced effective operators
$ \ocal_{\phi X} $ for $ X = W^I,\, B$ in (\ref{eq:lg.ops}).
In order to display explicitly the loop nature of these operators and including the fact that
gauge bosons couple universally, we will write
\beq
\qwe_{\phi W} = \frac{g^2}{16\pi^2} \tilde\qwe_W,
\qquad
\qwe_{\phi B} = \frac{g'{}^2}{16\pi^2} \tilde\qwe_B,
\eeq
so that these contributions to the effective Lagrangian become
\beq
\lcal_{\rm eff-loop}\up{\gamma\gamma} = \inv{\Lambda^2}
\left( \qwe_{\phi W} \ocal_{\phi W} + \qwe_{\phi B} \ocal_{\phi B} \right)
 = \frac \epsilon v \frac\alpha{4\pi} \tilde\qwe_{\gamma\gamma} \, \half H F_{\mu\nu} F^{\mu\nu},
\eeq
where $ \tilde\qwe_{\gamma\gamma} = \tilde\qwe_W + \tilde\qwe_B$.
Using the standard expressions for the top 
and $W$ loop contributions \cite{Okun:1982ap},  we find
\beq
\Gamma(H \to \gamma\gamma)
= \kappa_{\gamma\gamma}^2 \Gamma_{SM}(H \to \gamma\gamma)\,; \ 
\kappa_{\gamma\gamma}^2 =
1 - \qwe_{\partial\phi} \epsilon + {0.30} \tilde\qwe_{\gamma\gamma}\epsilon
+ {0.28}  \qwe_{t\vp}\epsilon.
\eeq

\item The $ H \to Z \gamma $ mode also receives contributions
from possible effective operator modifications of the $ H t t $ and
$HWW$ vertices,  as well as from $ \ocal_{\phi X} ~ X=W^I,\, B$
and $ \ocal_{WB} $ in (\ref{eq:lg.ops}) that generate
\beq
\lcal\up{Z\gamma}_{\rm 1 loop} = \frac{ e g}{16\pi^2}
\frac v{\Lambda^2} \tilde\qwe_{Z\gamma} F_\mn Z^\mn,
\eeq
where
\beq
\tilde\qwe_{Z\gamma} = \frac{16\pi^2}{eg}
\left[ \half( \qwe_{\phi W} - \qwe_{\phi B} ) s_{\rm 2w}
- \qwe_{WB} c_{\rm2w} \right],
\eeq
where $s_{\rm 2w}$ ($c_{\rm 2w}$) denotes the sine
(cosine) of twice the weak-mixing angle.
Using the known expressions for the loop factors
we find
\beq
\Gamma(H \to Z\gamma)
\!=\! \kappa_{Z\gamma}^2 \Gamma_{SM}(H \to Z\gamma)\,; \ 
\kappa_{Z\gamma}^2 \!=\!
1 - \qwe_{\partial\phi} \epsilon + {1.82} \tilde\qwe_{Z\gamma}\epsilon
+ {1.46} \qwe_{t\vp}\epsilon.
\eeq

\item Finally, the $H\to GG $ mode receives contributions
from $ \ocal_{t\vp}$ as well as from $ \ocal_{\phi G }$
in (\ref{eq:lg.ops}). Writing the coefficient of the latter
as $ \qwe_{\phi G} = g_s^2 \tilde\qwe_{GG}/(16\pi^2) $,
where $g_s$ is the $\su3_{\rm color}$ gauge coupling constant,
\beq
\Gamma(H \to GG)
= \kappa_{GG}^2 \Gamma_{SM}(H \to GG)\,; \ 
\kappa_{GG}^2 =
1 - \qwe_{\partial\phi} \epsilon + 2.91 \tilde\qwe_{GG}\epsilon
+ 4 \qwe_{t\vp}\epsilon.
\eeq
This mode can potentially receive significant radiative corrections,
however, explicit evaluation show that these are large only
for $ m_H > 2 m_t $ \cite{Spira:1995rr}, which is not the case.
(Radiative corrections to the lighter quark modes {\it are} large,
however, all contributions to the width from light quarks
are suppressed by
a factor $(m_q/v)^2 $, and can be ignored.)

\een

\subsection{Branching ratios and production cross section}

For $ m_H \sim 125~\gev $, the main decays of the SM Higgs 
are into the $b\bar b$ (58\%) and $WW^*$ (21\%) channels, 
but the $GG$ (9\%), $\tau\tau$ (6\%), $cc$ (3\%) and $ZZ^*$ (3\%)
are also significant
\beq
\Gamma(H) = \kappa_H^2 \Gamma_{SM}(H)\,; \qquad
\kappa_H^2 = 1- \qwe_{\partial\phi} \epsilon  + \beta \epsilon,
\eeq
where, using the \sm\ values available at~\cite{sm.br},
\bea
\beta &=&
\sum_{\psi=b,c,\tau} \qwe_{\psi\vp} \frac{\sqrt{2}\, v B(H \to \bar\psi\psi)}{m_\psi}+
(2.91 \tilde\qwe_{GG} + 4 \qwe_{t\vp})B(H \to \bar GG) \cr
&=& 43.115 \qwe_{b\vp} + 7.947 \qwe_{c\vp} + 12.385 \qwe_{\tau\vp}
+0.343 \qwe_{t\vp} + 0.249 \tilde\qwe_{GG}.
\eea
Then, for any final state $\xi,$ the branching ratio $B$ is related to the SM ratio $B_{SM}$ as
\beq
B(H \to \xi) = \frac{ \kappa^2_\xi}{\kappa^2_H} B_{SM}(H\to\xi).
\eeq 
Specifically,
\vskip-7mm
\bea
B(H \to \bar\psi\psi ) &=& \left( 1  + \frac{\sqrt{2}\, v}{m_\psi} \qwe_{\psi\vp}\epsilon-
\beta  \epsilon 
\right)B_{SM}(H \to \bar\psi\psi ) ,\cr
B(H\to Z\ell\ell) &=& 
\left( 1  - \beta \epsilon 
\right)B_{SM}(H\to Z\ell\ell),\cr
B(H\to W\ell\nu) &=& 
\left( 1 - \beta  \epsilon \right)
B_{SM}(H\to W\ell\nu),\cr
B(H\to \gamma\gamma) &=& 
\left( 1 + 0.30 \tilde\qwe_{\gamma\gamma}\epsilon
+ 0.28 \qwe_{t\vp}\epsilon - \beta  \epsilon \right)
B_{SM}(H\to \gamma\gamma) ,\cr
B(H\to Z\gamma) &=& 
\left( 1 + 1.82 \tilde\qwe_{Z\gamma}\epsilon
+ 1.46 \qwe_{t\vp}\epsilon - \beta  \epsilon \right)
B_{SM}(H\to Z\gamma) ,\cr
B(H\to GG) &=& 
\left( 1 + 2.91 \tilde\qwe_{GG}\epsilon
+ 4\qwe_{t\vp}\epsilon - \beta  \epsilon \right)
B_{SM}(H\to GG) .
\eea
Note that the ratio $ B(H\to Z\ell\ell)/B(H\to W\ell\nu) $ is expected to have deviations below 1\% from the SM value. Should this prove not to be the case, it would provide another strong indication of the presence of other light particles that affect these decays.

Although the decay mode $ H \to  GG$ has not been measured, the main contributions to the production cross section is in fact the inverse process
of gluon fusion.  Hence, to a good approximation we have
\beq
\sigma^{\rm prod} \simeq \kappa_{GG}^2 \sigma^{\rm prod}_{SM}.
\eeq
That might be probed, although it suffers from the usual difficulties
in determining the absolute normalization of a cross section.

The above new physics corrections are of order $ \epsilon $, which for
$ \Lambda > 1~\tev $ is smaller than $ 0.1 $. This provides a measure
of the precision that LHC experiments will need to reach in
order to probe physics at this scale (though sometimes a 
precision of 10\% might suffice, depending on how large the coefficients $\qwe_i$ and the numerical coefficients multiplying them are). None of the relevant experiments have (yet) reached this level, in fact, current experimental precision in these decays allows only the exclusion of new physics at scales that have already been probed directly.

%%%%%%%%%%%%%%%%%%%%
\section{Conclusion}\label{sec:conclude}

The purpose of this paper was to extend the analysis of ref.~\cite{Arzt:1994gp} to include couplings of the SM Higgs, especially those relevant to LHC measurements, in a {\it model-independent}\/  fashion, taking into account the existing constraints on breaking of the custodial $SU(2)$ symmetry of the Higgs sector of the SM. Implications for specific branching ratios were given in terms of the coefficients of various dimension-six operators.
The expressions presented show that any deviation of
the couplings $g_{HWW}$ or $g_{HZZ}$ from the SM, at a level of accuracy observable by LHC experiments ($\sim 10\%$ or higher), can be explained only by having $ \epsilon \sim 1 $, which corresponds to new physics at a scale below $ 1~\tev $. If the ATLAS  enhancement in the $ \gamma\gamma $ mode is verified, then this is strengthened considerably; it seems as if only new physics
around the electroweak scale could account for this effect.  Thus, we have sharpened the contrast between models that modify the SM by introduction of some higher mass scale and those, for example, having more than one Higgs doublet contributing to the weak scale \vev\ $v$; the current anomaly in the photon mode would then belong to the second possibility. As the data improve, it will be exciting to observe how BSM physics is first manifested.  We hope this analysis provides another tool by which this conclusion may be hastened.

\section*{Acknowledgement}
The research of one of us (MBE) was supported in part by the National Science 
Foundation under Grant No. NSF PHY11-25915.

\pagebreak

\appendix

%%%%%%%%%
\section{\!\! Dimension-Six Basis Operators for the SM\protect\footnote{These tables are taken from~\cite{Grzadkowski:2010es}, by permission of the authors. We changed the operator names from $Q$ to $\ocal$ to conform to the present conventions.}.} \label{sec:tables}

%%%%%%%%%
\vskip-5mm
\begin{table}[h]
\renewcommand{\arraystretch}{1.5}
\begin{tabular}{||c|c||c|c||c|c||}
\hline \hline
\multicolumn{2}{||c||}{$X^3$\ \bf{(LG)} } & 
\multicolumn{2}{|c||}{$\vp^6$~ and~ $\vp^4 D^2$ \bf{(PTG)} } &
\multicolumn{2}{|c||}{$\psi^2\vp^3$\ \bf{(PTG)} }\\
\hline
$\ocal_G$                & $f^{ABC} G_\mu^{A\nu} G_\nu^{B\rho} G_\rho^{C\mu} $ &  
$\ocal_\vp$       & $(\vp^\dag\vp)^3$ &
$\ocal_{e\vp}$           & $(\vp^\dag \vp)(\bar l_p e_r \vp)$\\
$\ocal_{\wt G}$          & $f^{ABC} \wt G_\mu^{A\nu} G_\nu^{B\rho} G_\rho^{C\mu} $ &   
$\ocal_{\vp\Box}$ & $(\vp^\dag \vp)\raisebox{-.5mm}{$\Box$}(\vp^\dag \vp)$ &
$\ocal_{u\vp}$           & $(\vp^\dag \vp)(\bar q_p u_r \tvp)$\\
$\ocal_W$                & $\eps^{IJK} W_\mu^{I\nu} W_\nu^{J\rho} W_\rho^{K\mu}$ &    
$\ocal_{\vp D}$   & $\left(\vp^\dag D^\mu\vp\right)^\star \left(\vp^\dag D_\mu\vp\right)$ &
$\ocal_{d\vp}$           & $(\vp^\dag \vp)(\bar q_p d_r \vp)$\\
$\ocal_{\wt W}$          & $\eps^{IJK} \wt W_\mu^{I\nu} W_\nu^{J\rho} W_\rho^{K\mu}$ &&&&\\    
\hline \hline
\multicolumn{2}{||c||}{$X^2\vp^2$\ \bf{(LG)}} &
\multicolumn{2}{|c||}{$\psi^2 X\vp$\ \bf{(LG)}} &
\multicolumn{2}{|c||}{$\psi^2\vp^2 D$\ \bf{(PTG)} }\\ 
\hline
$\ocal_{\vp G}$     & $\vp^\dag \vp\, G^A_{\mu\nu} G^{A\mu\nu}$ & 
$\ocal_{eW}$               & $(\bar l_p \sigma^{\mu\nu} e_r) \tau^I \vp W_{\mu\nu}^I$ &
$\ocal_{\vp l}^{(1)}$      & $(\vpj)(\bar l_p \gamma^\mu l_r)$\\
$\ocal_{\vp\wt G}$         & $\vp^\dag \vp\, \wt G^A_{\mu\nu} G^{A\mu\nu}$ &  
$\ocal_{eB}$        & $(\bar l_p \sigma^{\mu\nu} e_r) \vp B_{\mu\nu}$ &
$\ocal_{\vp l}^{(3)}$      & $(\vpjt)(\bar l_p \tau^I \gamma^\mu l_r)$\\
$\ocal_{\vp W}$     & $\vp^\dag \vp\, W^I_{\mu\nu} W^{I\mu\nu}$ & 
$\ocal_{uG}$        & $(\bar q_p \sigma^{\mu\nu} T^A u_r) \tvp\, G_{\mu\nu}^A$ &
$\ocal_{\vp e}$            & $(\vpj)(\bar e_p \gamma^\mu e_r)$\\
$\ocal_{\vp\wt W}$         & $\vp^\dag \vp\, \wt W^I_{\mu\nu} W^{I\mu\nu}$ &
$\ocal_{uW}$               & $(\bar q_p \sigma^{\mu\nu} u_r) \tau^I \tvp\, W_{\mu\nu}^I$ &
$\ocal_{\vp q}^{(1)}$      & $(\vpj)(\bar q_p \gamma^\mu q_r)$\\
$\ocal_{\vp B}$     & $ \vp^\dag \vp\, B_{\mu\nu} B^{\mu\nu}$ &
$\ocal_{uB}$        & $(\bar q_p \sigma^{\mu\nu} u_r) \tvp\, B_{\mu\nu}$&
$\ocal_{\vp q}^{(3)}$      & $(\vpjt)(\bar q_p \tau^I \gamma^\mu q_r)$\\
$\ocal_{\vp\wt B}$         & $\vp^\dag \vp\, \wt B_{\mu\nu} B^{\mu\nu}$ &
$\ocal_{dG}$        & $(\bar q_p \sigma^{\mu\nu} T^A d_r) \vp\, G_{\mu\nu}^A$ & 
$\ocal_{\vp u}$            & $(\vpj)(\bar u_p \gamma^\mu u_r)$\\
$\ocal_{\vp WB}$     & $ \vp^\dag \tau^I \vp\, W^I_{\mu\nu} B^{\mu\nu}$ &
$\ocal_{dW}$               & $(\bar q_p \sigma^{\mu\nu} d_r) \tau^I \vp\, W_{\mu\nu}^I$ &
$\ocal_{\vp d}$            & $(\vpj)(\bar d_p \gamma^\mu d_r)$\\
$\ocal_{\vp\wt WB}$ & $\vp^\dag \tau^I \vp\, \wt W^I_{\mu\nu} B^{\mu\nu}$ &
$\ocal_{dB}$        & $(\bar q_p \sigma^{\mu\nu} d_r) \vp\, B_{\mu\nu}$ &
$\ocal_{\vp u d}$   & $i(\tvp^\dag D_\mu \vp)(\bar u_p \gamma^\mu d_r)$\\
\hline \hline
\end{tabular}
\caption{\sf Dimension-six operators other than the four-fermion ones.\label{tab:no4ferm}}
\end{table}
\pagebreak
\begin{table}[h]
\begin{center}
{\bf All are PTG.}
\end{center}
\renewcommand{\arraystretch}{1.5}
\begin{tabular}{||c|c||c|c||c|c||}
\hline\hline
\multicolumn{2}{||c||}{$(\bar LL)(\bar LL)$} & 
\multicolumn{2}{|c||}{$(\bar RR)(\bar RR)$} &
\multicolumn{2}{|c||}{$(\bar LL)(\bar RR)$}\\
\hline
$\ocal_{ll}$        & $(\bar l_p \gamma_\mu l_r)(\bar l_s \gamma^\mu l_t)$ &
$\ocal_{ee}$               & $(\bar e_p \gamma_\mu e_r)(\bar e_s \gamma^\mu e_t)$ &
$\ocal_{le}$               & $(\bar l_p \gamma_\mu l_r)(\bar e_s \gamma^\mu e_t)$ \\
$\ocal_{qq}^{(1)}$  & $(\bar q_p \gamma_\mu q_r)(\bar q_s \gamma^\mu q_t)$ &
$\ocal_{uu}$        & $(\bar u_p \gamma_\mu u_r)(\bar u_s \gamma^\mu u_t)$ &
$\ocal_{lu}$               & $(\bar l_p \gamma_\mu l_r)(\bar u_s \gamma^\mu u_t)$ \\
$\ocal_{qq}^{(3)}$  & $(\bar q_p \gamma_\mu \tau^I q_r)(\bar q_s \gamma^\mu \tau^I q_t)$ &
$\ocal_{dd}$        & $(\bar d_p \gamma_\mu d_r)(\bar d_s \gamma^\mu d_t)$ &
$\ocal_{ld}$               & $(\bar l_p \gamma_\mu l_r)(\bar d_s \gamma^\mu d_t)$ \\
$\ocal_{lq}^{(1)}$                & $(\bar l_p \gamma_\mu l_r)(\bar q_s \gamma^\mu q_t)$ &
$\ocal_{eu}$                      & $(\bar e_p \gamma_\mu e_r)(\bar u_s \gamma^\mu u_t)$ &
$\ocal_{qe}$               & $(\bar q_p \gamma_\mu q_r)(\bar e_s \gamma^\mu e_t)$ \\
$\ocal_{lq}^{(3)}$                & $(\bar l_p \gamma_\mu \tau^I l_r)(\bar q_s \gamma^\mu \tau^I q_t)$ &
$\ocal_{ed}$                      & $(\bar e_p \gamma_\mu e_r)(\bar d_s\gamma^\mu d_t)$ &
$\ocal_{qu}^{(1)}$         & $(\bar q_p \gamma_\mu q_r)(\bar u_s \gamma^\mu u_t)$ \\ 
&& 
$\ocal_{ud}^{(1)}$                & $(\bar u_p \gamma_\mu u_r)(\bar d_s \gamma^\mu d_t)$ &
$\ocal_{qu}^{(8)}$         & $(\bar q_p \gamma_\mu T^A q_r)(\bar u_s \gamma^\mu T^A u_t)$ \\ 
&& 
$\ocal_{ud}^{(8)}$                & $(\bar u_p \gamma_\mu T^A u_r)(\bar d_s \gamma^\mu T^A d_t)$ &
$\ocal_{qd}^{(1)}$ & $(\bar q_p \gamma_\mu q_r)(\bar d_s \gamma^\mu d_t)$ \\
&&&&
$\ocal_{qd}^{(8)}$ & $(\bar q_p \gamma_\mu T^A q_r)(\bar d_s \gamma^\mu T^A d_t)$\\
\hline\hline 
\multicolumn{2}{||c||}{$(\bar LR)(\bar RL)$ and $(\bar L R)(\bar L R)$}\\
\cline{1-2}
$\ocal_{ledq}$ & $(\bar l_p^j e_r)(\bar d_s q_t^j)$ \\
$\ocal_{quqd}^{(1)}$ & $(\bar q_p^j u_r) \eps_{jk} (\bar q_s^k d_t)$ \\
$\ocal_{quqd}^{(8)}$ & $(\bar q_p^j T^A u_r) \eps_{jk} (\bar q_s^k T^A d_t)$ \\
$\ocal_{lequ}^{(1)}$ & $(\bar l_p^j e_r) \eps_{jk} (\bar q_s^k u_t)$ \\
$\ocal_{lequ}^{(3)}$ & $(\bar l_p^j \sigma_{\mu\nu} e_r) \eps_{jk} (\bar q_s^k \sigma^{\mu\nu} u_t)$ \\
\cline{1-2}
\cline{1-2}
\end{tabular}
\caption{\sf Four-fermion operators conserving baryon number. \label{tab:4ferm}}
\end{table}
\pagebreak

\end{document}